# Superconductivity in cubic noncentrosymmetric PdBiSe Crystal

**Bhanu Joshi, A Thamizhavel and S Ramakrishnan**

Department of Condensed Matter Physics and Materials Science
Tata Institute of Fundamental Research, Colaba, Mumbai – 400005, India

email: bpjphy@gmail.com

**Abstract**. Mixing of spin singlet and spin triplet superconducting pairing state is expected in noncentrosymmetric superconductors (NCS) due to the inherent presence of Rashba-type antisymmetric spin-orbit coupling. Unlike low symmetry (tetragonal or monoclinic) NCS, parity is isotropicaly broken in space for cubic NCS and can additionally lead to the coexistence of magnetic and superconducting state under certain conditions. Motivated with such enriched possibility of unconventional superconducting phases in cubic NCS we are reporting successful formation of single crystalline cubic noncentrosymmetric PdBiSe with lattice parameter a = 6.4316 Å and space group P2_1 3 (space group no. 198) which undergoes to superconducting transition state below 1.8 K as measured by electrical transport and AC susceptibility measurements. Significant strength of Rashba-type antisymmetric spin-orbit coupling can be expected for PdBiSe due to the presence of high Z (atomic number) elements consequently making it potential candidate for unconventional superconductivity.

## 1. INTRODUCTION

Unconventional superconductors has always been a source of fascination for research as they always adds a new understanding as well as challenges to demystify the long sought riddle for complete understanding of superconductivity. Noncentrosymmetryic Superconductors (NCS) are currently gaining lot of attention as a potential candidates to host rare coexistence of spin singlet and spin triplet superconducting pairing states and also due to the possibility of gapless non trivial topological phases in NCS and subsequent realization of Majorana fermions at NCS vortex core. Noncentrosymmetric superconductors lack the inversion symmetry in their underlying crystal structure which creates an antisymmetric spin-orbit coupling (ASOC) e.g. Rashba-type ASOC for tetragonal noncentrosymmetric crystal structure and depending upon its strength, it can affect superconducting as well as normal state properties of the system [1, 2, 3, 4, 5, 6, 7]. Contrary to the common belief, spin-triplet pairing is not completely excluded in such systems rather with sufficient strength this ASOC can induce mixing [8, 9, 10] of spin-singlet and spin-triplet pair components in the superconducting state by lifting the spin degeneracy of electrons and modifying the spin pattern in each sub band. NCS containing high Z elements are always preferred because strength of ASOC ∝ $Z^2$ and electronic wave vector **k**. Discovery of NCS CePt$_3$Si induced the birth of many NCS and despite many theoretical predications only very few viz. LiPt$_3$B, UIr and CeTX3 [8] have shown signatures of unconventional superconductivity. One of the prominent reasons for the absence of unconventional pairing could be a weak strength of ASOC in many of these compounds. Second important factor is the sample quality, structural disorder of the sample has a detrimental effect on the spin-triplet pairing. Considering these points in mind, our group in 2011 [11] has successfully discovered a new high quality (RRR~ 160), high Z element containing NCS: α-BiPd (monoclinic, Space group P2$_1$), which is now gradually turning out to be an important NCS.

Excited and motivated with the success of α-BiPd, in this present work we have formed cubic noncentrosymmetric PdBiSe single crystals (contains high Z elements) and established superconductivity in it via transport and magnetic measurements. Unlike low symmetry (tetragonal or monoclinic) NCS, parity is isotropicaly broken in the space for cubic NCS and can additionally lead to the coexistence of magnetic and superconducting state [12] under certain conditions.

## 2. EXPERIMENTAL DETAILS

The crystal was made by control heating of the individual components (Bi, 99.999% pure, Pd, 99.99% pure and Se 99.999%) in a high-purity Alumina crucible with a pointed bottom, which is kept in a quartz tube that is sealed under a vacuum of $10^{-6}$ mbar. Initially, the contents were heated up to 600 C in 24 h and then kept at 650 C for 24 h for the proper homogenization of the melt. Thereafter, it was slow cooled to 525 C with a rate of 1C/h and, finally, we come to room temp with a rate of 30 C/h. We obtained bulk shiny single crystals of PdBiSe with varying maximum dimension from 2 mm to 7 mm. The phase purity of the sample has been confirmed by powder x-ray diffraction measurement using Cu Kα radiation in a commercial diffractometer and subsequent Rietveld analysis [13] confirms the formation of PdBiSe having cubic structure with a noncentrosymmetric spacegroup P2_1 3 ( space group no 198) with lattice parameter a = 6.4316 (± 0.0001) Å. The structure consists of 12 atoms (4 formula unit) in the unit cell. Stoichiometry of the grown single crystals have also confirmed the by performing an energy dispersive analysis by x-ray (EDAX).

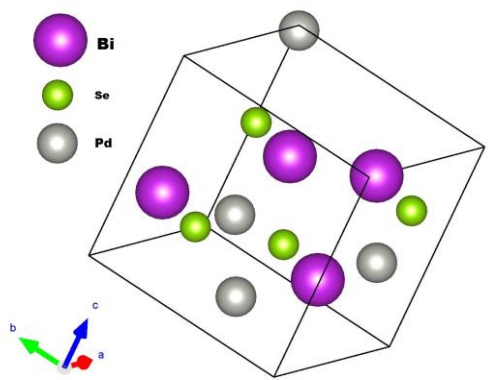

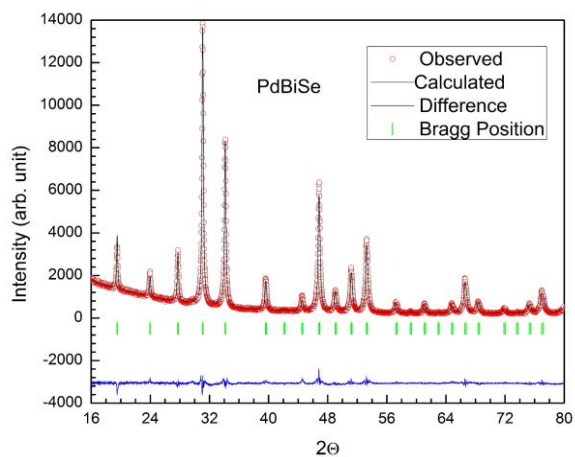

**Figure 1.** PdBiSe has cubic structure with a noncentrosymmetric spacegroup P2_1 3 and consists of 12 atoms (4 formula unit) in the unit cell. Crystal structure has been drawn using VESTA [14].

**Figure 2.** Powder x-ray diffraction data of the cubic NCS PdBiSe. The solid line is the simulated data using FullProf (Rietveld program).

Atomic position parameters extracted from Powder x-ray diffraction data of the cubic NCS PdBiSe are provided below in Table 1.

**Table 1. Atomic position parameters for unit cell of PdBiSe.**

|   | Ion | Name | x | y | z | Occ. | B | Site | Symmetry |
|---|-----|------|---|---|---|------|---|------|----------|
| 1 | Pd | Pd | 0.98485 | 0.98485 | 0.98485 | 1.000 | 1.000 | 4a | 0.3 |
| 2 | Bi | Bi | 0.62700 | 0.62700 | 0.62700 | 1.000 | 1.000 | 4a | 0.3 |
| 3 | Se | Se | 0.37639 | 0.37639 | 0.37639 | 1.000 | 1.000 | 4a | 0.3 |

## 3. RESULTS AND DISCUSSIONS

### 3.1. Transport and AC susceptibility studies

To characterize the superconducting transition one needs to do the transport, susceptibility and specific heat capacity measurements. Home-made set up of our lab is being used for transport measurement from 300 K down to 1.5 K. Sample contacts have been made using Silver paste and Gold wires (40 micron diameter) for the standard four-probe technique configuration to measure resistivity. Resistivity was measured using LR-700 (Linear Research, USA) bridge with 5 mA current having small AC frequency of 16 Hz. Figure 3 shows the temperature dependence of the resistivity for current orthogonal to (110) direction [$\rho(T)$] from 1.5 to 300 K. The good quality of the sample is clearly evident from the large residual resistivity ratio ($\rho_{300K}/\rho_{4K}$) of 16. The inset of Fig.3 shows superconductivity (Zero resistance) below 1.8 K. AC susceptibility (Figure 4) for PdBiSe crystal has been measured from 5 K down to 500 mK with field parallel to 110 direction and it reveals the superconducting transition below 1.5 K which is although at lower side of temperature but sharper in width as compared from transport.

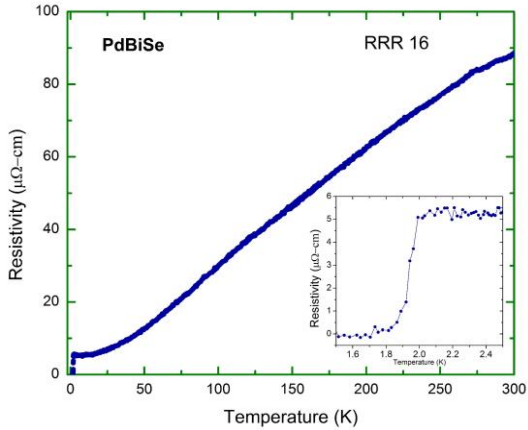
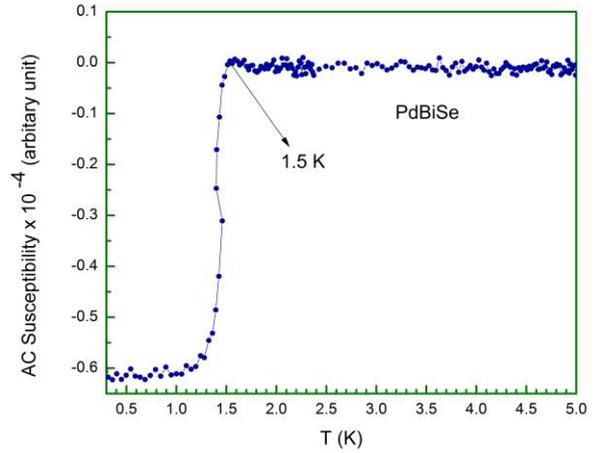

**Figure 3.** Temperature dependence of the resistivity for current orthogonal to (110) direction [$\rho(T)$] from 1.5 to 300 K.

**Figure 4.** AC susceptibility studies of PdBiSe revealed a SC transition below 1.5 K.

*3.2. Normal state specific heat capacity studies:*

Specific heat capacity ($C_p$) measurements has been performed from 300 K down to 1.8 K and no superconducting transition has been detected in this temperature range. The temperature dependence of Cp (Figure 5) is fitted to the expression $\frac{C_P}{T} = \gamma + \beta T^2$

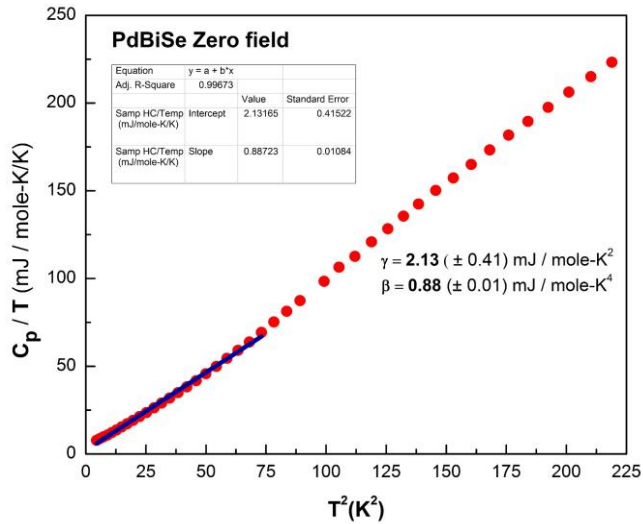

**Figure 5.** The low temperature (1.8 K to 15 K) part of Cp / T vs $T^2$ data in zero field.

Where γ better known as Somerfield coefficient is due to the electronic contribution and β is due to the lattice contribution. Fit yielded 2.13 (± 0.41) mJ/mol-$K^2$ and 0.88 (± 0.01) mJ/mol-$K^4$ for γ and β respectively, thus revealing PdBiSe to be a Non heavy fermion NCS.

**4. Conclusion**

A new cubic NCS PdBiSe has been produced and from best of our knowledge first time ever large single crystals of PdBiSe has been successfully grown. Signature of superconductivity in PdBiSe below 1.8 K are clearly found via transport and AC susceptibility measurement. Normal state heat capacity measurements are indicating non heavy fermion nature of PdBiSe. Mili-Kelvin heat capacity measurements are essential to establish bulk superconductivity in PdBiSe and they are in progress.